\documentclass[twocolumn,showpacs,showkeys,preprintnumbers,amsmath,amssymb]{revtex4}

\usepackage{graphicx}
\usepackage{dcolumn}
\usepackage{bm}

\begin{document}

\preprint{}

\title{Implementation of quantum search algorithm using classical Fourier optics}

\author{N. Bhattacharya, H. B. van Linden van den Heuvell and R. J. C. Spreeuw}
\email{spreeuw@science.uva.nl}
\affiliation{%
Van der Waals-Zeeman Institute, University of Amsterdam, \\
Valckenierstraat 65, 1018 XE Amsterdam, The Netherlands
}%
\homepage{http://www.science.uva.nl/research/aplp/}

\date{\today}

\begin{abstract}
We report on an experiment on Grover's quantum search algorithm showing
that {\em classical waves} can search a $N$-item database as efficiently
as quantum mechanics can. The transverse beam profile of a short laser
pulse is processed iteratively as the pulse bounces back and forth
between two mirrors. We directly observe the sought item being found in
$\sim\sqrt{N}$ iterations, in the form of a growing intensity peak on
this profile. Although the lack of quantum entanglement limits the {\em size}
of our database, our results show that entanglement is neither necessary for
the algorithm itself, nor for its efficiency.
\end{abstract}

\pacs{03.67.Lx, 42.25.-p, 42.30.Kq}

\keywords{Quantum searching, Fourier optics}
\maketitle


Quantum computers \cite{Fey82,Deu85} hold the promise of performing
tasks \cite{Sho97,Gro97} that are either impossible or much less
efficient without the use of quantum mechanics. One such task is quantum
searching \cite{Gro97,Gro97a}, introduced by Grover in a paper entitled
``{\em Quantum mechanics helps in searching for a needle in a
haystack}''. Consider using a phone book with $N$ entries to find the
name of a person whose phone number you have. Classically, this would
require $\sim N$ consultations of the phone book. Grover's algorithm
finds the desired entry with only $\sim\sqrt{N}$ consultations, using
quantum mechanics. Here we show experimentally that {\em classical
waves} can find a needle in a haystack as efficiently as quantum
mechanics can. Although some previous experiments
\cite{JonMosHan98,ChuGerKub98,KwiMitSch00,AhnWeiBuc00,DorLonAnd01} have
demonstrated various aspects of quantum searching, all of them have been
limited to four entries \cite{JonMosHan98,ChuGerKub98,KwiMitSch00} or a
single query \cite{KwiMitSch00,AhnWeiBuc00,DorLonAnd01}. Our experiment
closely follows Grover's algorithm, implementing for the first time an
iterative search on a 32-item database, using classical waves. It
provides a striking demonstration that the algorithm itself requires
only wave properties \cite{Llo00} but no entanglement \cite{Mey00}.

In Grover's (first) algorithm \cite{Gro97} each database item is
associated with a quantum state. Initially the system is prepared in a
superposition of all $N$ quantum states. The algorithm then amplifies
the probability amplitude of the state being sought, in an iterative
way. The item has been found once the probability amplitude of this
``target state'' is near unity. Ideally this requires $(\pi/4)\sqrt{N}$
iterations of the following two steps. In the first step a so-called
``oracle'' marks the item by inverting the phase of the associated
quantum state \cite{Gro97a}. In the second step the amplitudes of all
states are inverted about the average amplitude (IAA operation),
converting phase information into amplitude information.

\begin{figure}
\includegraphics{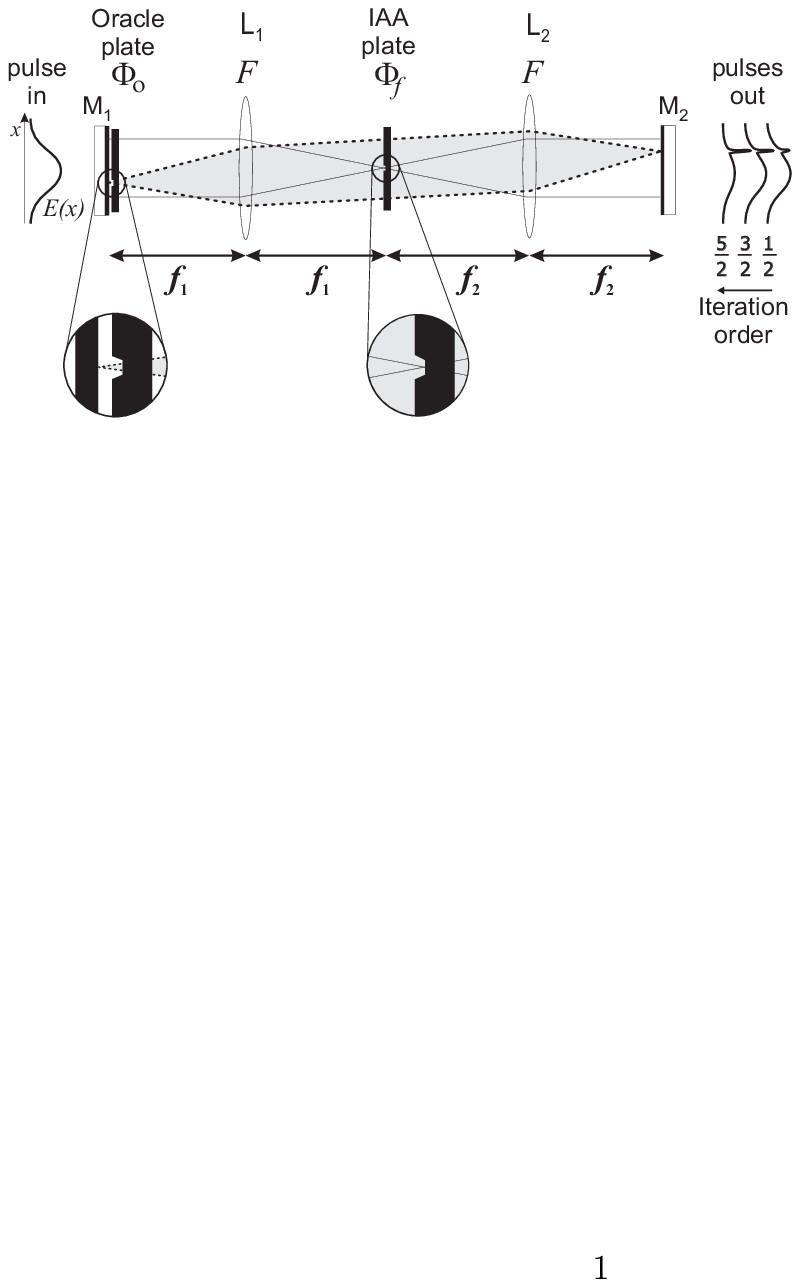}
\caption{\label{fig:cavity}Cavity implementing Grover's
algorithm using optical interference. We launch a short laser pulse with
a gaussian transverse beam profile $E(x)$, $x$ representing the data
register, into the cavity formed by mirrors M$_{1,2}$. A line shaped
depression in the oracle plate marks the item by imprinting a phase
profile $\Phi_o(x)$. The sequence $F\Phi_f FF\Phi_f F$ performs the
inversion about average (IAA) as required by Grover's algorithm. Here
$F$ denotes a Fourier transform, performed by the lenses L$_{1,2}$
(focal lengths $f_1 = 400$~mm, $f_2 = 600$~mm). The IAA plate imprints a
phase profile $\Phi_f(x')$ in the Fourier plane of the oracle. The
enlargements show cuts of the phase plates perpendicular to the lines.
As the pulse bounces back and forth, the transverse beam profile is
processed iteratively and light is concentrated into the shaded mode. A
high intensity peak, growing on the beam profile in the output plane,
indicates the sought item.} 
\end{figure}

The above protocol maps onto our classical-wave experiment as follows
(see Fig.~\ref{fig:cavity}). A complex electric field amplitude $E(x)$,
{\em viz.} a transverse laser beam profile plays the role of the quantum
probability amplitudes. The continuous coordinate $x$ labels the items
of the database, corresponding to all possible quantum states. By
spatial filtering we initialize the beam profile $|E(x)|^2$ as a smooth,
near-gaussian, distribution with a 1.33 mm diameter (FWHM; full width at
half maximum). A single, $\sim$ 300 ps laser pulse (wavelength 532 nm)
enters a standing-wave cavity of 2.02 m optical path length through
input mirror M$_1$ (transmission 2$\%$). The pulse travels back and
forth between the cavity mirrors in 13.5 ns, each roundtrip representing
one iteration of the search algorithm. Inside the cavity an ``oracle
plate'' \cite{Gro97a} marks the item by imprinting a phase profile on the
beam, $E(x) \rightarrow E(x)\exp(i\Phi_{o}(x))$, where $\Phi_{o}(x) =
\phi$ in a narrow area around the ``item position'' $x_{o}$ and
$\Phi_{o}(x) = 0$ elsewhere. Next, the IAA operation is performed by the
sequence $F\Phi_{f}FF\Phi_{f}F$, where $F$ denotes a Fourier transform
and $\Phi_{f}$ denotes a phase plate like the oracle, but now imprinting
a phase profile $\Phi_{f}(x^{\prime})$ in the Fourier plane. The Fourier
transforms replace the Walsh-Hadamard transforms \cite{Gro98} in the
original proposal \cite{Gro97} and are experimentally performed by
spherical, achromatic doublet lenses \cite{SalTei91}. Since $F^2$ is a
spatial inversion and $\Phi_{f}(x^{\prime}) = \Phi_{f}(-x^{\prime})$,
the IAA operation reduces to $F^{-1}\Phi_{f}^{2}F$. Thus the amplitude
amplifying Grover iterator is $\Phi_{o}^{2}F^{-1}\Phi_{f}^{2}F$. Note
that $F\Phi_{f}F$ can be recognized as phase contrast imaging.

\begin{figure}
\includegraphics{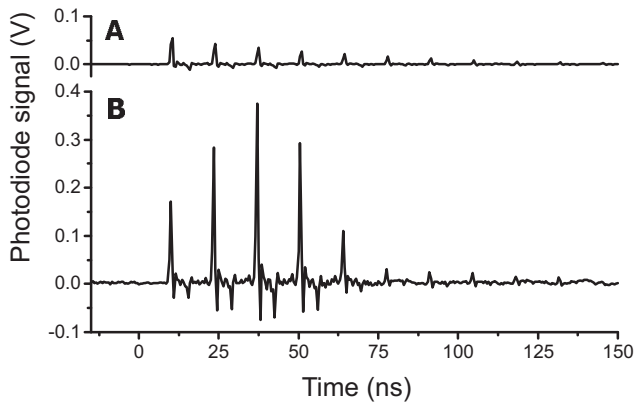}
\caption{\label{fig:ttrace} Amplitude amplification as observed in
trains of light pulses. Trace A shows the pulse train coupled out of a
bare cavity, i.e. with the phase lines shifted out of the beam. The peak
amplitudes decay exponentially with 25$\%$ roundtrip loss. For trace B the
phase lines were moved into the beam and the pulses were recorded behind
a narrow slit placed in the image of the oracle phase line. The energy
in the pulses increases, even though the total energy decays.}
\end{figure}

We observe the progress of the search algorithm iteration by iteration,
using the 2$\%$ transmission of mirror M$_{2}$ after each cavity
roundtrip. This light is imaged onto a 55 $\mu$m wide movable slit and
the transmitted light is collected on a photodiode. The photodiode
signal is amplified and recorded by a digitizing oscilloscope. The light
pulses are short compared to the roundtrip time, so that a train of
output pulses is obtained, one pulse per iteration. In
Fig.~\ref{fig:ttrace} we show two typical time traces. The trace in
Fig.~\ref{fig:ttrace}.A has been recorded in an ``empty cavity'',
leaving the oracle and IAA plates inside the cavity, but moving the
phase-shifting lines on the plates out of the beam. We observe an
exponentially decaying peak amplitude, with a roundtrip loss of about
0.25, due to reflections. Next, we move the IAA phase line into the beam
focus, put the oracle line in an arbitrary position in the beam, and
place the detection slit in the image of the oracle line. We then
observe a peak amplitude that {\em grows} during the first few
iterations, even though the total optical energy decreases. This is
shown in Fig.~\ref{fig:ttrace}.B and is a direct observation of
amplitude amplification.

We have measured the entire beam profile by recording traces like in
Fig.~\ref{fig:ttrace}.B for many different detection slit positions. We
combined the peak values at the same time from different traces into a
transverse beam profile. A sequence of such profiles for consecutive
roundtrips shows how the algorithm proceeds. In Fig. 3.A-C we show three
such sequences for increasing widths of the oracle line. Consecutive
profiles within a sequence have been multiplied by a factor $0.75^{-1}$,
in order to compensate for optical losses. We clearly observe the
solution growing as a high intensity peak in the transverse beam
profile. The position of this peak is the position $x_{o}$ of the sought
item, i.e. the phase line in the oracle, imaged by the intracavity
telescope. In the quantum case it would of course be impossible to watch
the solution grow as the algorithm proceeds, because a measurement would
cause the wave function to collapse.

On the basis of Grover's algorithm we expect the peak height to reach a
maximum after $(\pi/4)\sqrt{N/m}$ roundtrips, where $m$ is the number of
marked items \cite{BoyBraHoy98,Gro98a}, and to oscillate through a
sequence of maxima and minima with a period of $(\pi/2)\sqrt{N/m}$. In
an ideal, loss-free system, these cycles of finding and ``unfinding''
would continue indefinitely. This period assumes that the phase shifts
$\phi$ have their ideal values. Since we use the plates in double pass
inside the cavity, this ideal value is $\pm\pi/2$, whereas our measured
value is $\phi = -1.1 \pm 0.2$ rad. This increases the optimum number of
iterations to $[\pi/(4 \sin 1.1)]\sqrt{N/m}$. Although $\phi$ may
deviate from $\pi/2$, a ``phase matching'' condition
\cite{LonLiZha99,Hoy00} requires that the two phase shifts of the oracle
and IAA plates are approximately equal.

The ratio $N/m$ can be interpreted as the size of the database for a
single item search. Alternatively, the same $N/m$ also describes a
search for $m$ adjacent items in a larger database of size $N$. The
maximum database size is determined by optical diffraction, which limits
the effective number of positions $x$ that can be resolved. For our
cavity with a numerical aperture NA$=0.03$, the limit on the resolution
is given by Rayleigh's criterion as $0.61\lambda/$NA$\approx$10~$\mu$m.
For our 1.33 mm input beam, the maximum database size is then $\sim$133.

\begin{figure*}
\includegraphics[width=100mm]{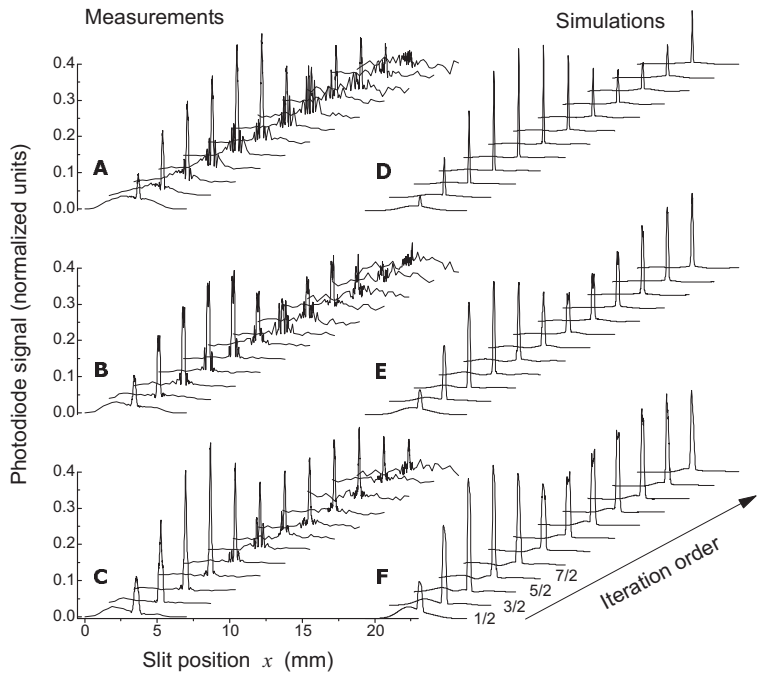}
\caption{\label{fig:xtrace} Iterative progress of the search algorithm
as shown by measured and simulated beam profiles. Oracle lines of width
(A) 42 $\mu$m, (B) 84 $\mu$m, and (C) 126 $\mu$m were used,
corresponding to databases of 31.7, 15.8, and 10.6 items, respectively.
Light coupled out of the cavity was recorded after each roundtrip
through a scanning slit. The peak growing in the first iterations
reveals the position of the sought item. The traces on the right (D, E,
F) have been simulated using realistic experimental parameters,
corresponding to those on the left.} \end{figure*}

We can estimate $N/m$ as the ratio of the input beam diameter to the
oracle line width. The phase shifting lines have been produced as the
shadows of thin metal wires (50, 100 and 200 $\mu$m diameter) while
evaporating a thin layer of SiO onto a BK7 substrate. A phase-contrast
image revealed line cross-sections that are well approximated by
trapezoids, with flat inner regions of 42, 84 and 126 $\mu$m, for the
oracle plate and 136 $\mu$m for the IAA plate. The deviations are probably
due to details of the evaporation procedure. Using the 1.33 mm diameter
(FWHM) of the input beam, we get expected ratios $N/m =$ 31.7, 15.8 and
10.6. We can compare this to the $N/m$ values as obtained from the
position of the first maximum in the search, bearing in mind that the
first image, having made $1/2$ roundtrip, should be counted as $1/2$
iteration. For the data shown in Fig.~\ref{fig:xtrace}.A-C we estimate
the maximum peak at 5, 3.5 and 3 iterations, leading to $N/m =$ 32,
15.8 and 11.6 respectively, in good agreement with the expected
numbers. The results thus confirm the $\sqrt{N/m}$ scaling behavior as
expected from Grover's algorithm.

The prime significance of the $N/m$ values is in the scaling of
the searching period as $\sqrt{N/m}$. The absolute values of the
expected $N/m$ depend on our chosen definition for the input beam
diameter (FWHM) and thus may seem somewhat arbitrary. Therefore we
also compare our results to a simulation, shown in
Fig.~\ref{fig:xtrace}.D-F. We simulate the phase plates by
multiplying the beam profile by the experimentally determined
trapezoidal phase profiles $\Phi_{o}(x)$ and $\Phi_{f}(x)$. We
describe the lenses by a Fourier transform. The results of the
simulation agree well with the experiment, producing the maximum
peak at the same number of iterations as the experiment. An
important difference between the experiment and the simulation is
due to optical losses in the experiment. As mentioned earlier, the
experimental data have been scaled to compensate for the losses,
which amplifies the noise in the last few iterations shown. Apart
from this noise, we also see the development of side peaks. These
are probably due to diffraction effects accumulating as the
iterations progress, e.g. due to slight misalignments of our
optical cavity.

Keeping the resolution at $\sim$10 $\mu$m and extending the experiment
to 2D, $E(x,y)$, it should be feasible to perform database searches of
up to $10^6$ items experimentally, assuming a beam diameter of 1 cm. This
is equivalent to about 20 qubits, so that we gain experimental access to
problems that are as yet inaccessible for true quantum computers. These
include quantum counting \cite{BoyBraHoy98,BraHoyTap98}, estimation of
the mean and median of a population \cite{Gro98a} and the synthesis of
arbitrary superposition states \cite{Gro00}. Theoretical studies have
investigated fault tolerance \cite{LonLiZha00,Son0010075} and noise
\cite{PabRui00} in Grover's algorithm, predicting damping of the cycles
of finding and ``unfinding'', like we also see in the experiment. The
problem of ``phase matching'' \cite{LonLiZha99,Hoy00} can also be
directly translated into optics as differential phase shifts provided by
the oracle and IAA plate. These issues are as yet impossible to
investigate experimentally with present-day quantum computers. Our
classical-wave experiment can bridge this gap. Note that it is
complementary to a theoretical proposal by Farhi and Gutmann
\cite{FarGut98} to search a digital database in analogue time, rather
than using discrete iterations. In our case, an analogue database is
searched using discrete iterations.

Some classical-wave analogies of quantum information processing
\cite{Spr98,CerAdaKwi98,Spr01}, as well as a hybrid quantum-classical
approach \cite{HowYea00} have been proposed previously. Some elements of
Grover's algorithm have been demonstrated with classical waves
\cite{KwiMitSch00}. The latter experiment demonstrated an oracle and IAA
operation for a four-item database. Iterations were neither present nor
necessary, since for $N = 4$ a single query reveals the sought item. A
four-item database search has also been demonstrated using NMR
techniques \cite{JonMosHan98,ChuGerKub98}. Electronic wave packets in
Rydberg atoms have been used to store and retrieve numbers
\cite{AhnWeiBuc00} and an equivalent experiment has been reported
recently with classical light waves \cite{DorLonAnd01}. However, it has
been pointed out that the Rydberg-atom experiment lacked the IAA
operation \cite{KwiHug00}, which is a crucial ingredient of the quantum
search algorithms. In our present experiment, Grover's second algorithm
\cite{Gro97a} can be recognized in the first transmitted pulse, which is
essentially a phase-contrast image of the oracle. Since the contrast
would be relatively low, the light pulse must contain sufficiently many
photons to build up good readout statistics. By contrast, using Grover's
first algorithm, the item could in principle be found with near
certainty by sending a single photon through the oracle $O(\sqrt{N})$
times.

It should be clear that our optical system is not a universal quantum
computer. Essentially we have mapped the $2^{n}$-dimensional Hilbert
space of $n$ qubits by the Hilbert space of a single photon in a
superposition of $2^{n}$ transverse modes. It is well known
\cite{Llo00,EkeJoz98} that this unary mapping comes at the cost of an
exponential overhead in some physical resource. Previous classical
analogies required an exponential number of components such as beam
splitters \cite{KwiMitSch00,Spr98,CerAdaKwi98,Spr01}. The efficiency of
a true quantum computer in implementing the transforms has been
attributed to entanglement, {\em i.e.} to the tensor product structure
of the Hilbert space. Despite the lack of entanglement in our present
experiment, the Fourier transform is performed efficiently using only a
single lens, independently of the size of the database. The lack of
entanglement does however limit the {\em size} of the database, which
scales linearly with the beam diameter $D$, or $\varpropto D^{2}$ for a
2D version. Thus the equivalent number of qubits scales only as
$\varpropto \log D$. Even if we set $D$ equal to the size of the
universe, $\sim 10^{26}$ m, this would yield only 206 equivalent qubits.
This limitation exists for any database containing classical
information. On the other hand, since Grover's algorithm provides only a
$\sqrt{N}$ speedup, a quantum computer implementing Grover's algorithm
becomes exponentially slow for an exponentially large database. Thus our
experiment shows that quantum entanglement is not needed to implement
the algorithm or to improve the efficiency. Its only role in this case
is to allow for a larger database size.

\begin{acknowledgments}
We would like to thank M. Groeneveld and H. Schlatter for the
preparation of the phase plates, G.H. Wegdam and T.W. Hijmans for use of
their equipment, and L.D. Noordam for a stimulating discussion. The
research of R.S. has been made possible by a fellowship of the Royal
Netherlands Academy of Arts and Sciences.
\end{acknowledgments}


\end{document}